\documentclass[twocolumn,amsmath,showkeys,prb]{revtex4-1}
\usepackage{dcolumn}
\usepackage{bm}
\usepackage[T1]{fontenc}
\usepackage{accents}
\usepackage{amsmath}
\usepackage{graphicx}
\usepackage{tabularx}
\usepackage{multirow}
\usepackage[usenames, dvipsnames]{color}
\usepackage{hyperref}
\usepackage{times}

\usepackage[normalem]{ulem}

\begin{document}

\title{Stone-Wales Defects in Hexagonal Boron Nitride as Ultraviolet Emitters}
\author{Hanen Hamdi}
\affiliation{Wigner Research Centre for Physics, P.O.\ Box 49, Budapest H-1525, Hungary}
\author{Gerg\H{o} Thiering}
\affiliation{Wigner Research Centre for Physics, P.O.\ Box 49, Budapest H-1525, Hungary}
\author{Zolt\'an Bodrog}
\affiliation{Wigner Research Centre for Physics, P.O.\ Box 49, Budapest H-1525, Hungary}
\author{Viktor Iv\'ady}
\affiliation{Wigner Research Centre for Physics, P.O.\ Box 49, Budapest H-1525, Hungary \\
Department of Physics, Chemistry and Biology, Link{\"o}ping University, SE-581 83, Link{\"o}ping, Sweden}
\author{Adam Gali}
\email{gali.adam@wigner.hu}
\affiliation{Wigner Research Centre for Physics, P.O.\ Box 49, Budapest H-1525, Hungary \\
Department of Atomic Physics, Budapest University of Technology and Economics, Budafoki \'ut 8, Budapest, H-1111, Hungary}
\date{\today}
\begin{abstract}
\textbf{Abstract}
Many quantum emitters have been measured close or near the grain boundaries of the two-dimensional hexagonal boron nitride where various Stone-Wales defects appear. We show by means of first principles density functional theory calculations that the pentagon-heptagon Stone-Wales defect is an ultraviolet emitter and its optical properties closely follow the characteristics of a 4.08-eV quantum emitter often observed in polycrystalline hexagonal boron nitride. We also show that the square-octagon Stone-Wales line defects are optically active in the ultraviolet region with varying gaps depending on their density in hexagonal boron nitride. Our results may introduce a paradigm shift in the identification of fluorescent centres in this material.
\end{abstract}

%\keywords{Hexagonal-Boron Nitride, two-dimensional material, Stone-Wales defect, formation energy,  (charged state) structural stabilities, zero phonon line, spin correlations}
\maketitle

\section{Introduction}

Two-dimensional (2D) materials are rich in novel phenomena in phyics. Hexagonal boron nitride ($h$-BN) is one of the key 2D materials in the field, which consists of boron-nitrogen bonds in a honeycomb lattice. Because of the strong polarization of the covalent bond between boron and nitrogen, it exhibits a large gap of about 6~eV~\cite{Cassabois2016} unlike the semimetallic graphite or graphene. Recently, room temperature quantum emitters have been found in $h$-BN~\cite{Tran2016, Bourrellier2016}. These findings have attracted large interest as they may serve as a basis to realize room temperature photon quantum blockade or very sensitive quantum sensors~\cite{Abdi2019}. In particular, the first quantum emitters were found to emit in the visible~\cite{Tran2016} and one in the ultraviolet (UV) region~\cite{Bourrellier2016}. The origin of these emitters is unknown but assumed to come from point defects with states in the fundamental band gap of $h$-BN~\cite{Tran2016}. It can be envisioned in 2D materials that these quantum emitters may be created in a well controlled fashion because the composition of the top layer can be directly manipulated with different techniques (e.g., Ref.~\onlinecite{Krivanek2010}), but first, the nature of these quantum emitters should be identified, in order to further develop quantum optics measurements on these quantum emitters and their deterministic creation.

We focus our attention to the UV quantum emitter~\cite{Bourrellier2016} that has a zero-phonon line (ZPL) energy at around 4.08~eV with prominent phonon sideband peaks~\cite{Museur2008, Bourrellier2016} (see Fig.~\label{fig:PL}). This defect has a Debye-Waller factor ($DW$), i.e., the ratio of ZPL intensity and the total intensity in the luminescence, at $\approx0.14$ which corresponds to $S\approx2$ Huang-Rhys factor. The optical lifetime of the emitter was observed at $\approx1$~ns~\cite{Museur2008}. We note that this colour centre is distinct from similar UV emitters~\cite{Vuong2016, Pelini2019} that have ZPL emission at around 4.1~eV but less pronounced phonon sideband with $S\approx1$ (see Discussion and Supplementary Note 1 for further discussion).

% main results in front
Here, we computed the optical properties of the pentagon-heptagon Stone-Wales defect by Kohn-Sham hybrid density functional theory in $h$-BN. We find that the calculated ZPL energy, the phonon modes participating in the fluorescence spectrum (see Fig.~\ref{fig:PL}), and the optical lifetime agree with the observed data on the 4.08-eV quantum emitter. The metastable triplet state has characteristic zero-field splitting due to the low symmetry of the defect where the corresponding spin states may be selectively addressed under illumination. Our study implies that abundant emitters may exist with relatively high formation energies near or inside grain boundaries of hexagonal boron nitride, and extended defects could be the origin of other colour centres in this material.
\begin{figure}[ht!]
 \includegraphics[width=.45\textwidth]{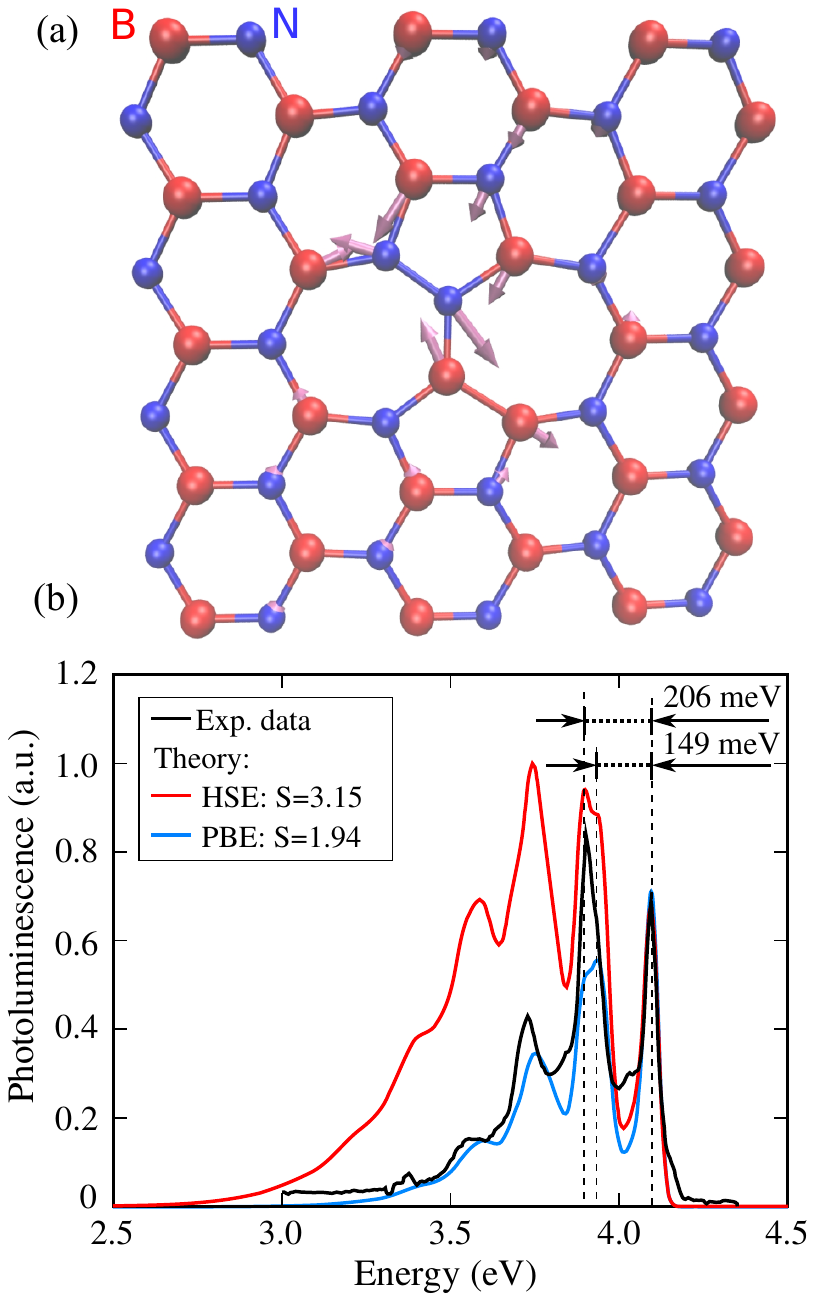}
 % vmdscene.png: 3000x2000 px, 72dpi, 105.83x70.56 cm, bb=0 0 3000 2000
 %\includegraphics[width=.45\textwidth]{PLS-modified.pdf}
 % luminiscence-spectrum.png: 1123x794 px, 96dpi, 29.71x21.01 cm, bb=0 0 842 595
 \caption{\label{fig:PL}Photoluminescence of the UV quantum emitter. (a) Structure of the pentagon-heptagon Stone-Wales defect in hexagonal boron nitride as optimized by HSE DFT (see Methods). The arrows represent the change in the atomic positions going from the optimized ground state to the optimized electronic excited state. (b) The experimental (excitation by 4.5-eV laser as reported in Ref.~\onlinecite{Museur2008}) (black curve) and calculated PL spectrum by HSE (red curve) and PBE (blue curve). The ZPL position of HSE was aligned by +0.02~eV whereas that of PBE was aligned by +0.65~eV, in order to directly compare the features in the phonon sideband, and Gaussian broadening of 20~meV was applied for the ZPL peak and the phonon sideband. The first prominent feature in the PL sideband nearest to ZPL contains two overlapping phonon modes that are shown with vertical lines with the corresponding phonon energies. S is the calculated Huang-Rhys factor.}
\end{figure}

\section{Results}
\subsection{Electronic structure and formation energy}
\begin{figure}[!ht]
\includegraphics[width=.5\textwidth]{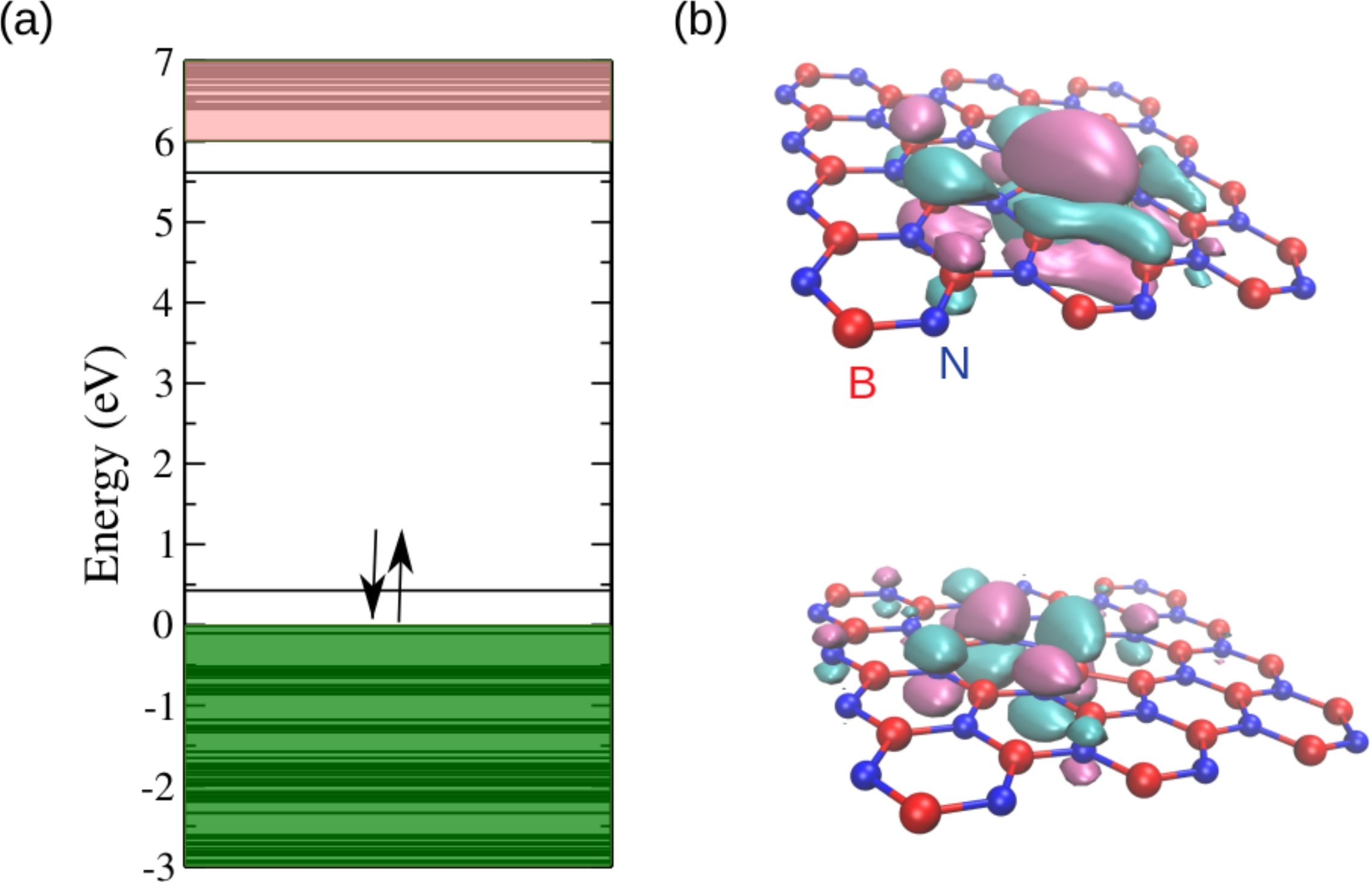}
\caption{\label{fig:lev_wfs}Electronic structure of the pentagon-heptagon Stone-Wales defect in hexagonal boron nitride. (a) HSE Kohn-Sham levels. (b) Corresponding defect wavefunctions for the empty state (top) and occupied state (bottom). The real space wave functions are visualized by cyan and mauve lobes representing the isosurface of the wave function at +0.0005 1/\AA$^{3}$ and -0.0005 1/\AA$^{3}$ values, respectively.}
\end{figure}

We apply first principles plane wave density functional theory (DFT) calculations on the pentagon-heptagon rings Stone-Wales defect in $h$-BN (see Methods).
The pentagon-heptagon structure in $h$-BN can be created by rotating one boron-nitrogen pair by 90 degrees in the hexagonal lattice about the axis perpendicular to the $h$-BN sheet which automatically introduces a nitrogen antisite and a boron antisite with creating a nitrogen-nitrogen bond and a boron-boron bond, respectively [see Fig.~\ref{fig:PL}(a)]. We find that the nitrogen-nitrogen bond creates a level at $E_\text{v}+0.44$~eV whereas the boron-boron bond creates a level at $E_\text{c}-0.40$~eV in the fundamental band gap, where $E_\text{v}$ and $E_\text{c}$ are the valence band maximum and conduction band minimum, respectively (see Fig.~\ref{fig:lev_wfs}). This structural defect is isovalent with the perfect lattice, thus the lower energy level is fully occupied whereas the upper level is empty with constituting a closed shell singlet electronic configuration. This is an electrically and optically active defect as the defect may be ionized and optical transition can occur between the occupied and empty defect levels in the gap. Indeed, previous scanning electron microscope measurements associated a 2.5~eV energy gap with the pentagon-heptagon Stone-Wales defect in $h$-BN~\cite{Li2015} but the calculated levels do not confirm this interpretation because the electronic gap rather remains in the UV region ($\approx5.2$~eV). 

The formation energy ($E^\text{form}(\text{SW}^q)$)  and adiabatic ionization energies of the defect can be calculated by the thermodynamic equation developed by Zhang and Northrup~\cite{Zhang1991} with using a charge correction energy ($\Delta^{q}$) with a charge $q$ of the defect from Refs.~\cite{Freysoldt2009, Freysoldt2018} as follows,
\begin{equation}
\label{eq:Ef}
E^\text{form}(\text{SW}^q) = E_\text{tot}(\text{SW}^q)-E_\text{tot}(\text{BN})+q(E_\text{v}+E_\text{F})+\Delta^{q}\text{,}
\end{equation}
where $E_\text{tot}(\text{SW}^q)$ and $E_\text{tot}(\text{BN})$ are the total energy of the defect in the charge state $q$ and the perfect supercell, respectively, whereas $E_\text{F}$ is the Fermi level between $E_\text{v}$ and $E_\text{c}$. The calculated formation energy of the neutral defect is 7.2~eV which is relatively high and it is basically in accord with a previous DFT PBE study~\cite{Wang2016}. The calculated donor and acceptor levels in the single sheet of $h$-BN are resonant with $E_\text{v}$ and $E_\text{c}$, respectively.

\subsection{Optical properties}
In the neutral defect, the optical transition may occur between the occupied and unoccupied defect states in the gap in the UV region. In the Franck-Condon approximation, the strength of optical transition does not change as a function of the coordinate of the ions and the participation of the phonons in the optical transition, i.e. the phonon sideband in the luminescence spectrum, can be calculated as the overlap between the phonon modes in the adiabatic potential energy surface (APES) of the electronic ground state and of the electronic excited state. We apply this theory to characterize the optical transition of this defect. We further simplify this picture by assuming that the APES in the electronic ground state and excited state is very similar, and thus the corresponding phonons are equivalent, i.e., we apply Huang-Rhys approximation. In the Huang-Rhys approximation, the calculated Huang-Rhys factor ($S$) has a direct relation to the observed Debye-Waller factor, $DW$ as $DW = \exp (-S)$. 

As a consequence, the calculation of the luminescence spectrum requires several steps: i) computing the electronic excited state with geometry optimization, ii) computing the phonons in the electronic ground state, iii) computing the overlap between the phonon modes in the electronic ground and excited states. We go through these steps one-by-one, with providing analysis on the results.

\subsubsection{Electronic excited state}
Before computing the electronic excited state, it is intriguing to analyze the electronic structure by group theory. The pentagon-heptagon Stone-Wales defect has a single mirror plane symmetry, $C_{1h}$. Both the lower energy and upper energy defect states show $a^{\prime\prime}$ symmetry because they are basically $p_z$ orbitals localized on the nitrogen and boron atoms around the nitrogen-nitrogen and boron-boron bonds, respectively. The closed shell singlet state is an ${}^1A^\prime$(g) state. Excited states can be constructed by promoting an electron from the lower defect level to the upper defect level in the gap. This can produce a ${}^3A^\prime$ triplet and a ${}^1A^\prime$(e) singlet excited state. The ${}^3A^\prime$ state is dark, and it is lower in energy than the  ${}^1A^\prime$(e) because of the exchange interaction of the two electrons. The many-body 
${}^1A^\prime$(e) is a correlated state that can be only described by two Slater-determinants. The spinpolarized hybrid DFT method might not be able to produce an accurate charge density of this interacting many-electron system, therefore the $\Delta$SCF method has larger inaccuracy than anticipated (e.g., $\approx0.1$~eV in energy, see Ref~\onlinecite{Gali2009}) for less severe excited states. This can affect the optimized geometry in the excited state and the total energy of the ${}^1A^\prime$(e) state. The latter can be corrected by estimating the exchange energy by the calculated DFT total energies of the singlet and triplet excited states similarly to a previous work~\cite{Mackoit2019}. At the ground state geometry, the corrected excitation energy is 4.53~eV which is reduced by 0.47~eV in the geometry optimization procedure. The final calculated ZPL energy is 4.06~eV which perfectly agrees with the ZPL energy of the 4.08-eV emitter. 

\subsubsection{Phonon density of states}
\begin{figure*}[ht!]
\includegraphics[width=.95\textwidth]{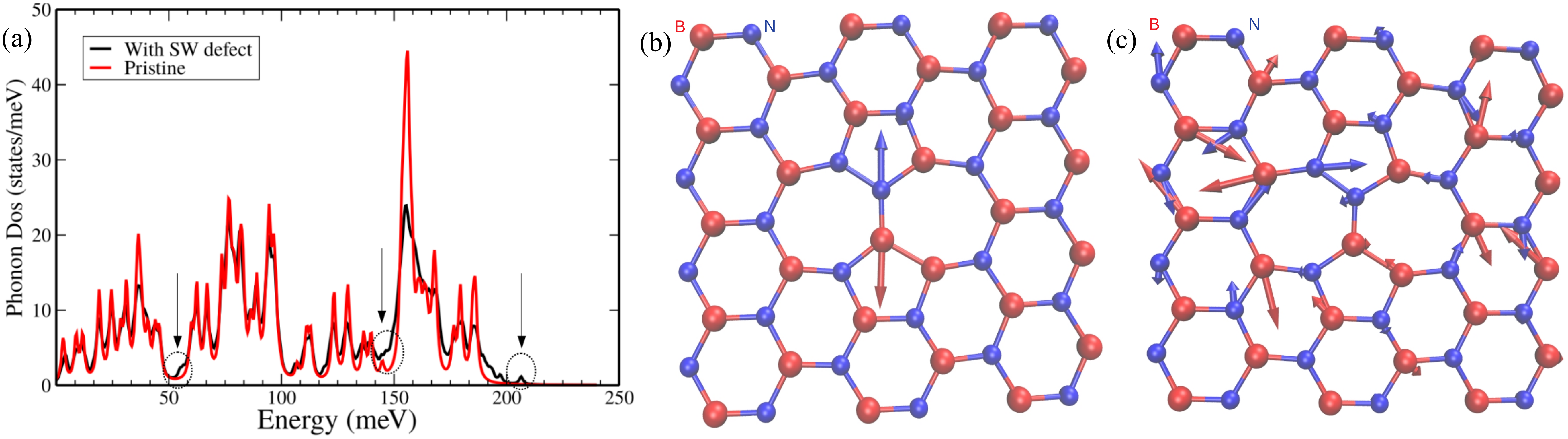}
%\includegraphics[width=.5\textwidth]{Phonon-Dos.png}
 %\includegraphics[width=.5\textwidth]{mode1-last.pdf}
 %\includegraphics[width=.5\textwidth]{mode-154-scale.pdf}
 % mode1.png: 7000x2000 px, 72dpi, 246.94x70.56 cm, bb=0 0 7000 2000
% phonoon-DOS.png: 605x340 px, 96dpi, 16.00x8.99 cm, bb=0 0 454 255
\caption{\label{fig:phonons}Phonons of the pentagon-heptagon Stone-Wales defect in hexagonal boron nitride. (a) Calculated phonon density of states for the pristine (red curve) and defective (black curve) hexagonal boron nitride. The position of the (quasi)local phonon modes are highlighted by circles and arrows. The motion of ions in the highest energy (quasi)local phonon modes at (b) 206~meV and (c) 149~meV. The length of the arrows represents the relative amplitude of the motion of the ions where larger scale was applied for (c) for the sake of visibility.}
\end{figure*}

The calculated phonon densities of states for the perfect and defective supercells are shown in Fig.~\ref{fig:phonons}(a). As can be seen, the defect introduces new phonon modes at about 206~meV slightly above the phonon bands, and two other ones at 149~meV and 54~meV, respectively. The two highest energy phonon modes asymmetrically stretch the boron-boron and nitrogen-nitrogen bonds, as reported in Fig.~\ref{fig:phonons}(b) and (c). In the optical excitation, these bonds are indeed stretched [see Fig.~\ref{fig:PL}(a)], therefore they should appear in the corresponding optical excitation spectrum.

\subsubsection{Computed photoluminescence spectrum and optical lifetime}
Next, we turn to the contribution of phonons to the optical transition. The computed luminescence spectrum is shown in Fig.~\ref{fig:PL}(b).
The calculated phonon prominent peaks in the phonon sideband occurs exactly at those energies that were observed in the experiment, and their corresponding replica can be also well recognized. The first intense peak near the ZPL line consists of predominantly two broad phonon modes at 149~meV and 206~meV that were analyzed above. On the other hand, the present calculational approach seems to overestimate the intensity of the phonon sideband and the calculated $S\approx3$ is larger than the observed one at $\sim2$ for the 4.08-eV emitter~\cite{Museur2008}. Since the optimized geometry of the ${}^1A^\prime$(e) state directly enters the calculation of the luminescence phonon sideband and the Huang-Rhys factor that suffers from an error due to the spin contamination it can cause a larger inaccuracy in the calculated intensity of the phonon sideband than anticipated for other defects~(e.g., Refs. \onlinecite{Thiering2017, Gali2019}). We note that our PBE calculations predict $S\approx2$ for this defect which may be due to a fortuituous cancellation of errors in the PBE functional for this particular state that finally provides a good optimized geometry for the electronic excited state. As a consequence, the calculated phonon sideband based on the PBE geometries provide excellent agreement with the observed PL phonon sideband. The main point of these findings is that the peak positions in the phonon sideband of the 4.1-eV emitter can be well-explained by the (quasi)localized phonon modes associated by the Stone-Wales defect. Our analysis emphasizes the importance of the local or quasilocal phonon modes in understanding the PL spectrum. The prominent phonon modes either coincide or are very close to the corresponding bulk phonon modes in energy, e.g., 200-meV Raman mode of bulk $h$-BN, but they are rather localized on the defect with producing relatively sharp phonon replicas in the luminescence spectrum. 
    
\begin{figure}[!ht]
\includegraphics[width=.5\textwidth]{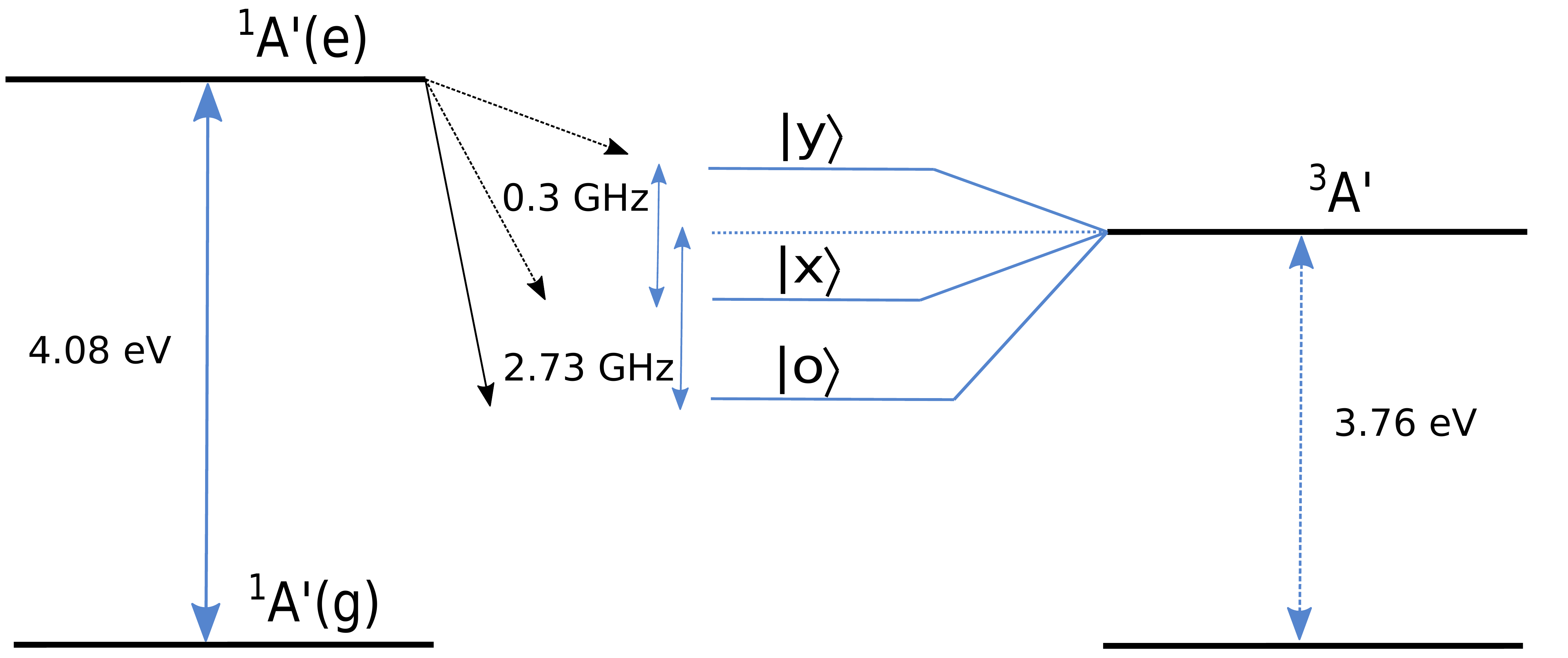}
 % excited -precess.png: 1123x794 px, 96dpi, 29.71x21.01 cm, bb=0 0 842 59
 % wave-function-ground-state-draw.pdf: 0x0 px, 300dpi, 0.00x0.00 cm, bb=
\caption{\label{fig:levels}The electronic states and levels of pentagon-heptagon Stone-Wales defect in hexagonal boron nitride. For the sake of clarity, the energies are not scaled. The zero-field splitting is depicted for the triplet state. Intersystem crossing can occur between ${}^1A^\prime$(e) and the $ms=0$ of ${}^3A^\prime$ state by spin-orbit interaction (solid line) whereas electron-phonon coupling weakly enables it towards the $ms=\{x,y\}$ states.}
\end{figure}

Finally, we address the optical lifetime of the defect which depends on the radiative and non-radiative rates. The non-radiative decay from the ${}^1A^\prime$(e) may occur via the intersystem crossing (ISC) towards the ${}^3A^\prime$ state. The ${}^3A^\prime$ triplet state splits via dipolar electron spin-spin interaction where the calculated zero-field splitting falls in the GHz region (see Fig.~\ref{fig:levels}). Group theory analysis implies that the ISC towards the $ms=0$ state is faster than that towards $ms=\{x,y\}$ states. The ISC rate can be calculated as (Refs.~\onlinecite{Goldman2015, Thiering2018, Gali2019}),
\begin{equation}
\label{eq:ISC}
\Gamma _\text{nr} = 4 \pi \hbar \lambda_z^2 F(\Delta) \text{,}
\end{equation}
where $\lambda_z=0.31$~GHz is the calculated spin-orbit coupling between ${}^1A^\prime$(e) and ${}^3A^\prime$, whereas $F$ is the phonon overlap spectral function between ${}^1A^\prime$(e) and ${}^3A^\prime$ separated by $\Delta=0.3$~eV, which yields very small value because $S\approx0.1$ between the two states. As a consequence, $\Gamma_\text{nr} \propto 0.1$~kHz. Thus, the non-radiative decay rate can be neglected in the optical lifetime which can be then calculated from the radiative lifetime as
\begin{equation}
\label{eq:lifetime}
   \tau = \frac{3\pi\epsilon_0\hbar c^3}{n\omega^3|\mu|^2} \text{,} 
%= \frac {3\pi^2\hbar c^3}{n\omega^3V}(\int Im \epsilon(E)dE)^{-1} = \frac {3\pi^2\hbar c^3}{n\omega^3V} A^{-1}
\end{equation} 
where $n=2.75$ is the refractive index of $h$-BN at $\hbar\omega=4.53$~eV vertical excitation energy~\cite{Cappellini2001}, $\mu$ is the optical transition dipole moment as derived from the calculated imaginary part of the dielectric function between the corresponding Kohn-Sham wavefunctions representing the ground and excited states at the ground state geometry, $\epsilon_0$ is the vacuum permittivity, $c$ is the speed of light. We obtain $\tau = 1.8$~ns, which is close to the observed one. 

We conclude from the calculated ZPL energy, characteristic phonon replica in the phonon sideband, and the optical lifetime that the 4.08-eV quantum emitter can be associated with the pentagon-heptagon Stone-Wales defect in $h$-BN.

\section{Discussion} \label{sec:discussion}
\subsection{Comparison to other models}
UV emission with the ZPL energy at around 4.1~eV was often found in polycrystalline $h$-BN, therefore these emitters were erronously associated with the free exciton of $h$-BN in the past~\cite{Solozhenko2001}. Other early experimental and theoretical studies assumed~\cite{Katzir1975, Era1981} that the UV emitters are associated with the presence of carbon defects because carbon is a common impurity during the growth of $h$-BN. Later, it was proposed based on some analog with the donor-acceptor pair emission in GaN and related materials, that the UV emission may include two types of defects with donor and acceptor nature involving a carbon-related defect~\cite{Du2015, Uddin2017}. However, the observed fast optical lifetime goes strongly against this argument and rather indicates a defect with localized electronic ground and excited states as alluded in Ref.~\onlinecite{Mackoit2019}.  Recently, a carbon-dimer defect has been proposed as the origin for the 4.1-eV emitters based on first principles calculations~\cite{Mackoit2019}.

The confusion about the origin of the UV emitters may be caused by the fact that these UV emitters could have very similar ZPL energies at around 4.1~eV but they belong to different colour centres. In a recent work~\cite{Pelini2019}, a series of UV emitters with ZPL energies at around 4.1~eV has been detected, including emitters with $S\approx1$ phonon sideband~\cite{Vuong2016}. They prepared ${}^{13}$C doped $h$-BN too and observed the same series of UV emitters but no isotope shift was found in the corresponding PL spectra~\cite{Pelini2019} (see Supplementary Note 1). Furthermore, one UV emitter with ZPL energy at 4.08~eV always appeared in the sample whereas the PL intensity of the other UV emitters with blue shift compared to the 4.08-eV emitter exibited some correlation with the carbon content of $h$-BN. Our study focuses on the single UV emitter~\cite{Bourrellier2016} with ZPL emission at 4.08~eV which has a broad prominent peak shifted from the ZPL energy between 150~meV and 200~meV, and it has a Huang-Rhys factor at $S\approx2$. This type of UV emitter was previously measured as ensembles in polycrystalline $h$-BN~\cite{Museur2008}. 

We realized that the 4.08-eV emitter may be situated near or inside the grain boundaries of polycrystalline $h$-BN. In the grain boundary regions, pentagon-heptagon structural Stone-Wales defects have been observed in the hexagonal lattice~\cite{Gibb2013}, a very well known type of defects in carbon nanotubes or graphene sheets. The Stone-Wales defect contains boron and nitrogen antisites in $h$-BN [see Fig.~\ref{fig:PL}(a)], therefore it has relatively high formation energy~\cite{Wang2016, Weston2018}; however, the large strain at the grain boundaries mediates the formation of Stone-Wales defects~\cite{Mortazavi2014}. Previous studies already confirmed~\cite{Liu2012, Li2015, Wang2016} that pentagon-heptagon Stone-Wales defects may introduce levels into the band gap but the feasible optical signature has not yet been explored. We find here that they produce the commonly observed UV emission. 

\subsection{4.08-eV optical centre as a quantum bit}
We note that the calculated very slow ISC rate implies that the $S=1$ state cannot be effectively pumped optically from the ground state. Continous wave photo-excitation might drive the system very slowly to the $S=1$ state but absorption of the UV photon in the $S=1$ state would immediately photo-ionize the pentagon-heptagon Stone-Wales defect into a dark state, thus the lifetime of the $S=1$ state under continuous UV illumination would be short. On the other hand, injection of holes and electrons into $h$-BN can result in trapping the exciton into the $S=1$ state by pentagon-heptagon Stone-Wales defects which should have a long lifetime in dark. We tentatively propose that the $S=1$ state plays a role in the $\sim$100~ns antibunching signal in the cathodoluminescence measurements of the 4.08-eV emitter~\cite{Meuret2015}. The relatively long lifetime of the $S=1$ state would make possible to induce quantum bit operation with alternating magnetic fields in the microwave region when the mobile bulk exciton is trapped into this state of the pentagon-heptagon Stone-Wales defect. 

\subsection{Structural defects and quantum emitters}

\begin{figure*}[!ht]
\includegraphics[width=.95\textwidth]{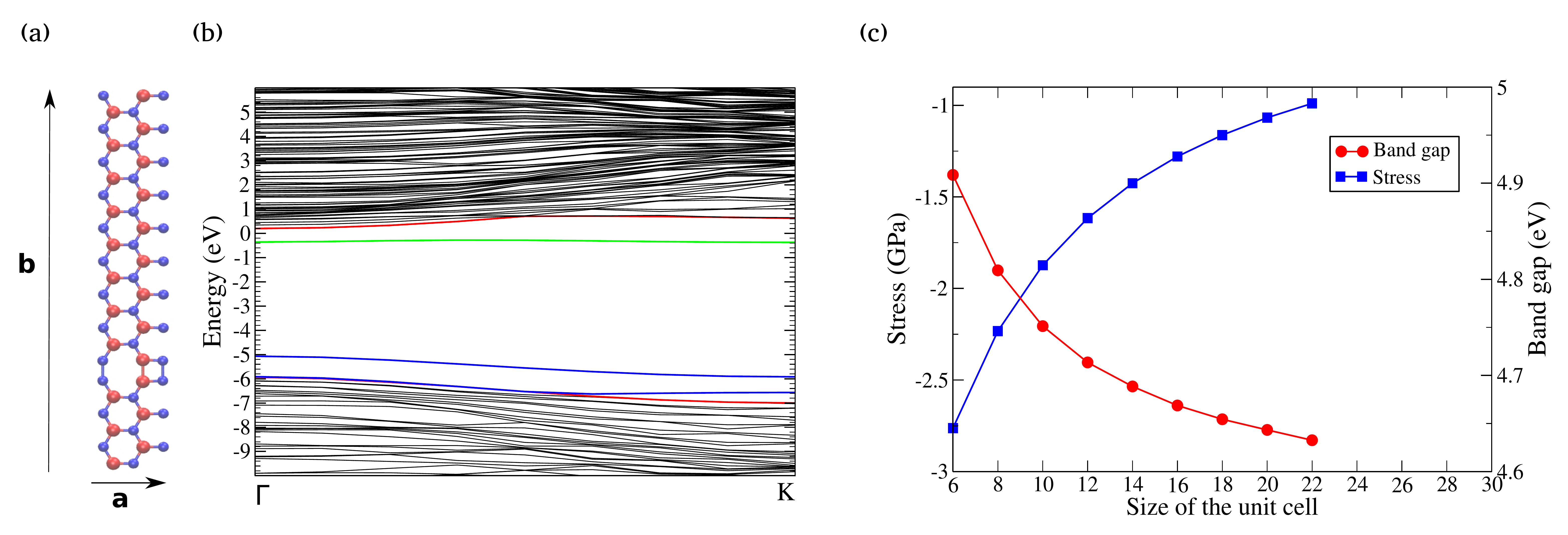}
\caption{\label{fig:48bands}The electronic bands and band gap of square-octagon Stone-Wales line defect in hexagonal boron nitride. (a) Structure of square-octagon Stone-Wales line defect (see Ref.~\onlinecite{Cretu2014}). The density of the defect was varied by adding pristine $h$-BN units in the $b$ direction with increasing the distance between the defects which releases the strain induced by the defect. The nitrogen atoms from the neighbour unit cell is depicted along $a$ direction for showing the square defect. (b) HSE band structure for $b=12$ defective lattice. Red curves are the valence band maximum (VBM) and conduction band minimum (CBM) of the pristine $h$-BN, whereas the other coloured bands are the defect related bands near the VBM and CBM. (c) Shift of the electronic band gap as a function of the distance between the line defets and self-induced stress between the defects as obtained by HSE. The stress occurs only in the sheet and no stress shows up in the direction normal to the $h$-BN sheet but the unit of stress is still given in GPa.}
\end{figure*}

Beside pentagon-heptagon Stone-Wales defect, square-octagon Stone-Wales defects can occur in $h$-BN at the grain boundaries~\cite{Liu2012}. Among the different types of square-octagon Stone-Wales defects~\cite{Liu2012, Cretu2014, Li2015}, scanning tunneling microscope studies clearly identified those structures that have N-N and B-B bonds (see Fig.~\ref{fig:48bands}) which often forms line defects~\cite{Cretu2014}. Here we only consider this type of square-octagon Stone-Wales defect. 
The square-octagon Stone-Wales line defect with N-N and B-B bonds create bands: the N-N occupied bands (blue curves) appear above and near the valence band maximum of the pristine $h$-BN whereas the B-B empty band (green curve) appears below the conduction band minimum of the pristine $h$-BN (see Fig.~\ref{fig:48bands}). The line defect related band structure is direct at the $\Gamma$-point in which the band gap strongly varies by the distance between the line defects because the strain field induced by them (see Fig.~\ref{fig:48bands}(c)). The band gap of isolated square-octagon Stone-Wales line defect approaches 4.6~eV. These line defects could also reside relatively close to each other~\cite{Cretu2014}, and then the band gap can reduce substantially in the order of 100~meV. Thanks to the direct gap nature, the line defect is optically active and may also contribute to the ultraviolet emission in $h$-BN.

In summary, we have reported the results of hybrid functional calculations for the pentagon-hexagon Stone-Wales defect in hexagonal boron nitride. The calculated optical properties of this Stone-Wales defect, in particular, the phonon modes participating in the optical transitions, reproduce all the known features of a 4.08-eV quantum emitter. Our study shows that polycrystalline $h$-BN may contain such defects that have relatively high formation energy but still abundant and optically active. 

Furthermore, the extended Stone-Wales line defects introduces occupied and empty bands in the fundamental band gap of pristine $h$-BN which effectively reduces the band gap for proximate point defects. As a consequence, the ionization energies of these point defects are reduced that can affect their photo-stability as found for divacancy defects inside stacking faults in 4H SiC~\cite{Ivady2019}. Moreover, some common defects in $h$-BN, like carbon substitutional defects~\cite{Weston2018}, have optical transition between the in-gap defect level and band edges. Since the effective band edges shift closer to the in-gap defect level for defects near the Stone-Wales line defect the corresponding optical signal will change with showing generally longer wavelength emission. Stone-Wales line defects also generate a strain field that can act on the optical transition of point defects, in which the optical transition occurs between in-gap defects states but sensitive to the strain field, e.g., vacancy complexes~\cite{Li2020}.

As many optical centers and quantum emitters were found near or inside the grain boundaries or cracks of $h$-BN this study may turn the direction of research towards exploring the magneto-optical properties of structural defects and the interaction between point defects and structural defects for identification of the observed colour centres in $h$-BN. 

\section{Methods}
\subsection{First principles methodology}
First principles calculations are performed using Kohn-Sham spinpolarized density functional theory (DFT) with the \textsc{vasp} package~\cite{Kresse1996}.
We use the screened hybrid functional of Heyd, Scuseria,
and Ernzerhof (HSE) \cite{doi:10.1063/1.1564060, doi:10.1063/1.2204597}. In this approach, the short-range exchange potential is calculated by mixing a fraction of
nonlocal Hartree-Fock exchange with the generalized gradient
approximation of Perdew, Burke, and Ernzerhof (PBE)~\cite{Perdew1996}.
The screening parameter is set to 0.2~\AA $^{-1}$ and the mixing parameter to $\alpha = 0.32$. These parameters closely reproduce the experimental band gap and structural parameters of $h$-BN, similarly to GaN and AlN materials \cite{Moses2011}. 
%A correction for the van der Waals interactions is included within the Grimme-D3 scheme \cite{Grimme2010}. 
The valence electrons are separated from the core electrons by projector augmented wave (PAW) potentials \cite{Blochl1994}. An energy cutoff of 500~eV is used for the plane-wave basis set. We model the pentagon-heptagon Stone-Wales (SW) defect in a 162-atom $h$-BN supercell in which $\Gamma$-point calculation for mapping the Brillouin-zone suffices. The vacuum size is set to 25~\AA . In the geometry optimization procedures, the force criterion is set to 0.001~eV/\AA . The phonons are determined by calculating the numerical derivatives of the forces for creating the Hessian matrix in the ground state electronic configuration within PBE. The spin-orbit coupling is calculated in non-collinear approximation with the spin axis set perpendicular to the $h$-BN sheet within HSE. The square-octagon line defect was calculated the same parameters but modelled in a rectangular unit cell of $h$-BN with using $4\times 1\times 1$ k-point mesh. We note that previous calculations on well-localized defect states showed that the single sheet model and the $h$-BN bulk model produce very similar ZPL energies for the same defect, also as a function of strain~\cite{Li2020}. We expect here the same behaviour for the considered defects. 

The magneto-optial properties are calculated with employing the \emph{ab initio} toolkit developed in our group~\cite{Gali2019}. Briefly, the excited state is calculated by the $\Delta$SCF method~\cite{Gali2009} where the correction for the total energy of the open-shell singlet excited state is applied, similarly to a previous work~\cite{Mackoit2019}. The phonon sideband in the calculated luminescence spectrum is calculated within the Huang-Rhys approximation using our home-built implementation~\cite{Thiering2017} in the spirit of Ref.~\onlinecite{Alkauskas2014}. The dipolar electron spin-spin interaction part of the zero-field splitting is calculated as implemented by Martijn Marsman (see also Ref.~\onlinecite{Bodrog2014}).

\section{Data availability}
The data that support the findings of this study are available from the authors on reasonable request, see author contributions for specific data sets.

\bibliography{biblio}

\section{Acknowledgements}
 
A.G.\ acknowledges the support from the National Office of Research, Development and Innovation in Hungary for Quantum Technology Program (Grant No.\ 2017-1.2.1-NKP-2017-00001), National Excellence Program (Grant No.\ KKP129866), and the EU H2020 Asteriqs project (Grant No.~820394). V.I.~acknowledges the MTA Premium Postdoctoral Research Program and the support from the Knut and Alice Wallenberg Foundation through WBSQD2 project (Grant No.\ 2018.0071). We acknowledge the fruitful discussions from Guillaume Cassabois and technical help from Joel Davidsson and B\'alint Somogyi.

\section{Author information}
\subsection{Contributions}
A.G.\ wrote the manuscript with input from all authors. H.H., V.I.\ and A.G.\ carried out calculations, Z.B.\ did the group theory analysis. G.T.\ developed the electron-phonon code and applied it together with H.H.\ and A.G. The computational results were analyzed with contributions from all authors. The research was initiated and supervised by A.G.

\subsection{Corresponding author}
Correspondence to Adam Gali.

\section{Ethics declarations}
\subsection{Competing interests}
The authors declare no competing interests.

\end{document}